\documentclass[prb,twocolumn,showpacs,preprintnumbers,amsmath,amssymb]{revtex4}
\usepackage{graphicx}

\begin{document}

\title{Effect of frustration on charge dynamics for a doped two-dimensional triangular Hubbard lattice: Comparison with a square lattice}

\author{T. Tohyama}
\email{tohyama@imr.tohoku.ac.jp}
\affiliation{Institute for Materials Research, Tohoku University, Sendai, 980-8577, Japan.}
\date{\today}

\begin{abstract}
We examine the optical conductivity $\sigma(\omega)$ and the chemical potential $\mu$, together with the spin correlation, in the strong-coupling limit of a hole-doped two-dimensional triangular Hubbard model near half filling by using an exact diagonalization technique. In contrast to the case of a square lattice without frustration, the doping dependences of $\mu$ and the Drude weight indicate that the charge degree of freedom is weakly coupled to the spin degree of freedom. However, we find that $\sigma(\omega)$ shows strong incoherent excitations extended to a higher energy region. This implies that geometrical frustration in strongly correlated electron systems influences incoherent charge dynamics.
\end{abstract}

%\keywords{}%Use show keys class option if keyword display desired

\maketitle

The interplay of geometrical frustration and strong electron correlation is one of the hot topics in the field of strongly correlated electron systems. The simplest example of such geometry is a triangular lattice in two dimensions (2D).~\cite{Ong}  

Numerical studies for a spin-1/2 triangular Heisenberg model have suggested the presence of the antiferromagnetic (AF) long-range order of a three-sublattice coplanar type~\cite{Elser,Bernu,Capriotti} in spite of frustration. However, the order is expected to be easily destroyed by carrier doping and a resonant-valance-bond (RVB) state~\cite{Anderson} may emerge. In fact, it has been suggested that for the positive sign of hopping ($t>0$) the RVB state is favored with hole doping~\cite{Koretsune} and $d+\mathrm{i}d$-wave superconductivity appears.~\cite{Watanabe}  It was also suggested that the three-sublattice magnetism is stable in a wide range of doping.~\cite{Weber} On the other hand, for $t<0$ the Nagaoka ferromagnetism~\cite{Nagaoka} emerges with doping.~\cite{Koretsune,Watanabe,Weber,Merino} These results indicate that geometrical frustration and strong correlation give rise to strong competition among many states.     

The variety of the ground states in the triangular lattice may induce unconventional electron excitations and charge dynamics through coupling with frustrated magnetism. However, there are few theoretical studies about charge dynamics away from half filling, except for the fermion-spin theory of the $t$-$J$ model.~\cite{Yu,Liu} Therefore, it is very important to investigate the effect of frustration on charge dynamics and to clarify the interplay of the charge and spin degrees of freedom in geometrically frustrated systems. For this purpose, comparison with a square lattice is very crucial, because it has no geometrical frustration.

In this paper, we perform an exact diagonalization study of the doping dependences of the optical conductivity $\sigma(\omega)$, the chemical potential $\mu$, and static spin correlation for a twenty-one-site triangular Hubbard cluster with large on-site Coulomb interaction. We introduce an averaging procedure over various twisted boundary conditions (BCs) to reduce finite-size effects. From spin correlation near half filling, we confirm strong three-sublattice magnetic correlation for $t>0$ and a tendency toward ferromagnetism for $t<0$, as reported before.~\cite{Koretsune,Watanabe,Weber,Merino} We find that $\mu$ and the Drude weight $D$ are roughly proportional to the hole concentration $\delta$, irrespective of the sign of $t$. These behaviors are similar to those of the one-dimensional (1D) Hubbard model where the charge and spin degrees of freedom are decoupled, but different from the case of the square lattice where the coupling is strong.  Therefore, charge in the triangular lattice is expected to be weakly coupled to spin. However, we find that $\sigma(\omega)$ shows very strong incoherent excitations extended to a high energy region of several $t$. Such  high-energy incoherent excitations are not seen in the square lattice. This implies that geometrical frustration causes unconventional charge dynamics.

The Hubbard model with the nearest-neighbor hopping $t$ and on-site Coulomb interaction $U$ is given by
\begin{equation}
H_\mathrm{Hub}= -t\sum_{\left<i,j\right>,\sigma} \left(c_{i,\sigma }^\dagger c_{j,\sigma}+\mathrm{H.c.}\right) +U\sum_i n_{i,\uparrow}n_{i,\downarrow},
\label{H_Hub}
\end{equation}
where $c_{i,\sigma}$ annihilates an electron with spin $\sigma$ at site $i$,  $n_{i,\sigma}=c_{i,\sigma }^\dagger c_{i,\sigma}$, and the summation $\left< i,j \right>$ runs over nearest-neighbor pairs.
Being interested in the region of $U\gg t$, we take the strong-coupling limit of Eq.~(\ref{H_Hub}). The resulting Hamiltonian reads $H_\mathrm{SC}=H_{tJ}+H_\mathrm{3S}$
with
\begin{equation*}
H_{tJ}= -t\sum_{\left<i,j \right>,\sigma}
\left( \tilde{c}_{i,\sigma }^\dagger \tilde{c}_{j,\sigma} + \mathrm{H.c.} \right) + J\sum_{\left<i,j\right>}\left( \mathbf{S}_i\cdot \mathbf{S}_j-\frac{1}{4}\tilde{n}_i\tilde{n}_j\right),
\label{H_t-J}
\end{equation*}
and
\begin{eqnarray*}
H_\mathrm{3S}&=&-\frac{J}{4} \sum_{\left<i,j,k\right>,\sigma,\sigma'} \left(1-n_{i,-\sigma}\right) c_{i,\sigma}^\dagger c_{j,\sigma} n_{j,-\sigma}\nonumber \\
&& \ \times n_{j,-\sigma'} c_{j,\sigma'}^\dagger c_{k,\sigma'} \left(1-n_{k,-\sigma'}\right),
\label{H_3S}
\end{eqnarray*}
where $J=4t^2/U$, $\tilde{c}_{i,\sigma}=c_{i,\sigma}(1-n_{i,-\sigma})$, $\tilde{n}_i=\sum_\sigma \tilde{c}_{i,\sigma}^\dagger \tilde{c}_{i,\sigma}$, and
$\left<i,j,k\right>$ denotes a pair of three nearest-neighbor sites. We use $U/t=20$, which is above a critical $U/t$ [$\simeq$ 12.1 (Ref.~13) or $\simeq$12.4 (Ref.~14] for the metal-insulator (MI) transition at half filling.

The exact diagonalization method based on the Lanczos algorithm is frequently applied to the Hubbard-type models as an unbiased numerical method. For the triangular lattice, one uses a $N$-site cluster with $\mathbf{R}_a=l\mathbf{u}+m\mathbf{v}$ and $\mathbf{R}_b=-m\mathbf{u}+(l+m)\mathbf{v}$, being that $N=l^2+m^2+lm$ with integers $l,m\ge 0$ and $\mathbf{u}$ and $\mathbf{v}$ are the vectors connecting nearest-neighbor sites given by $\mathbf{u}=\mathbf{x}$ and $\mathbf{v}=\frac{1}{2}\mathbf{x}+\frac{\sqrt{3}}{2}\mathbf{y}$ with the unit vector $\mathbf{x}$ ($\mathbf{y}$) in the $x$ ($y$) direction. In this study, we take $N=21$ $(l=4, m=1)$ and introduce carriers up to five holes. In such a small cluster, we are not free from finite-size effects. In order to reduce the finite-size effects, we introduce various BC with twist and average physical quantities over the twisted BC. This procedure has been applied for various quantities in $t$-$J$-type models.~\cite{Poilblanc,Tohyama04} The twist induces the condition that $c_{i+R_a,\sigma}=e^{\mathrm{i}\phi_a}c_{i,\sigma}$ and $c_{i+R_b,\sigma}=e^{\mathrm{i}\phi_b}c_{i,\sigma}$ with arbitrary phases $\phi_a$ and $\phi_b$.  Note that $\phi_a=\phi_b=0$ corresponds to periodic BC.
$\phi_{a(b)}$ is defined as $\phi_{a(b)}=\boldsymbol{\kappa}\cdot\mathbf{R}_{a(b)}$
with $\boldsymbol{\kappa}=\kappa_x\mathbf{x}+\kappa_y\mathbf{y}$.
$\boldsymbol{\kappa}$ is taken within an area surrounded by four corners at  $(\kappa_x,\kappa_y)=\pm\frac{\pi}{N}(l,\frac{1}{\sqrt{3}}(l+2m))$ and $\pm\frac{\pi}{N}(l+2m,-\sqrt{3}l)$. For the averaging procedure, we choose many $\boldsymbol{\kappa}$ with equal intervals of $\pi/40$ in the area. The total number of $\boldsymbol{\kappa}$ results in $N_\kappa=353$, except for the five-hole case where we take $N_\kappa=$177 because of time-consuming calculations. For the square lattice, we use a $N=20$ cluster~\cite{Tohyama04} and $N_\kappa=320$.
We note that, at half filling ($\delta$=0) where the Hamiltonian is given by the Heisenberg model, the averaging procedure can not be applied because of the absence of hopping terms.

\begin{figure}
\begin{center}
\includegraphics[width=8.cm]{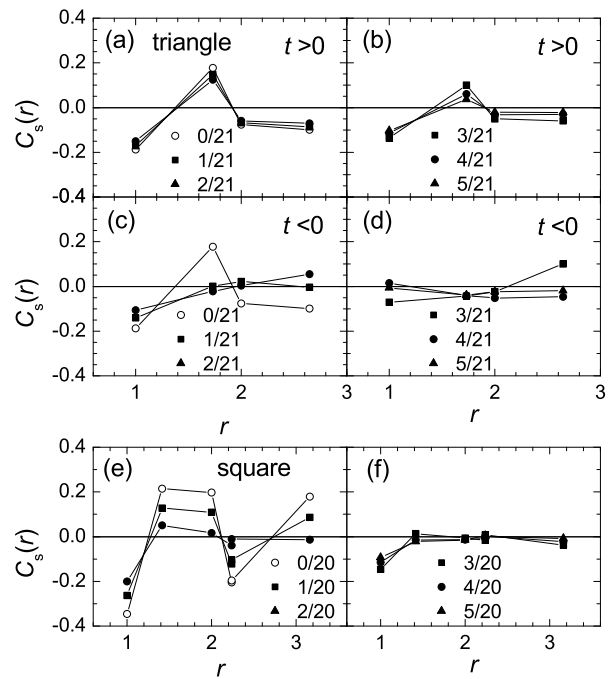}
%\vskip4cm
\caption{\label{fig1}
Spin correlation $C_\mathrm{s}$ as a function of two-spin distance $r$  for various $\delta$ in the strong-coupling limit of the Hubbard model with $U/t=20$. (a),(b) the $N=21$ $t>0$ triangular lattice. (c), (d) the $N=21$ $t<0$ triangular lattice. (e),(f) the $N=20$ square lattice. The numbers in the panels represent $\delta$.}
%\vspace{6mm}
\end{center}
\end{figure}

We first examine the static spin correlation function $C_\mathrm{s}(r)$ between two sites separated by a distance $r$. Using the averaging procedure, we define this as
\begin{equation*}
C_\mathrm{s}(r)=\frac{1}{N_\kappa}\sum_{\boldsymbol{\kappa}} \frac{1}{NN_r} \sum_{\mathbf{i},\boldsymbol{\rho}} \left< \Psi^{\boldsymbol{\kappa}}_0 \left| \mathbf{S}_{\mathbf{i}+\boldsymbol{\rho}}\cdot\mathbf{S}_\mathbf{i} \right|\Psi^{\boldsymbol{\kappa}}_0\right>,
\label{Cspin}
\end{equation*}
where $\left| \Psi^{\boldsymbol{\kappa}}_0\right\rangle$ represents the ground state with energy $E^{\boldsymbol{\kappa}}_0$ for a $\boldsymbol{\kappa}$, the summation of $\boldsymbol{\rho}$ runs over bonds satisfying $\left|\boldsymbol{\rho}\right|=r$, and $N_r$ is the number of the bonds. Figure~\ref{fig1} shows the doping dependence of $C_\mathrm{s}(r)$. In the triangular lattice with $t>0$ [Figs.~\ref{fig1}(a) and \ref{fig1}(b)], $C_\mathrm{s}(r)$ for $\delta>0$ shows $r$ dependences similar to that for $\delta=0$. This behavior is consistent with a previous report~\cite{Weber} that the three-sublattice magnetism at half filling is stable in a wide range of $\delta$. We also find that this doping dependence is different from the case of the square lattice shown in Figs.~\ref{fig1}(e) and \ref{fig1}(f), where AF spin correlation at $\delta=0$ is rapidly suppressed with doping because of the strong coupling of charge motion and spin background. Therefore, this comparison may tempt us to have a view that the spin and charge degrees of freedom are weakly coupled in the $t>0$ triangular lattice. However, such a view is different from the spin-charge separation seen in 1D systems, as will be discussed below in terms of $\sigma(\omega)$. The $t<0$ triangular lattice shows a different behavior.  Even for single-hole doping,  $C_\mathrm{s}(r)$ dramatically changes from that of the Heisenberg model as shown in Fig.~\ref{fig1}(c), and approaches to small values toward $\delta=5/21$. This may be a precursor of ferromagnetic ground states.~\cite{Koretsune,Watanabe,Weber,Merino} In fact, the total spin at $\delta=5/21$ under the averaging procedure is $S_\mathrm{ave}=0.441$. We note that a fully polarized ferromagnetic ground state is obtained at $\delta=7/21$ under periodic BC. It is interesting to notice that the doping dependence of $C_\mathrm{s}(r)$ is small except for the dramatic change from zero to one hole. This is also different from the case of the square lattice, where the change from zero to three holes is continuous. The small doping dependence for $t<0$ also seems to be indicative of weak coupling of spin and charge.

In small clusters, the chemical potential $\mu$ at a given $\delta$ can be expressed as 
$\mu=\left[e_0(\delta_1)-e_0(\delta_2)\right]/(\delta_2-\delta_1)$,
where $e_0(\delta)$ is the ground-state energy per site evaluated by $e_0(\delta)=(N_\kappa N)^{-1}\sum_{\boldsymbol{\kappa}} E_0^{\boldsymbol{\kappa}}$ and $\delta_1$ and $\delta_2$ satisfy $\delta=(\delta_1+\delta_2)/2$.
Here we note that the averaging for $e_0(\delta)$ is crucial for obtaining reliable doping dependence of $\mu$ because the averaging reduces the finite-size effects and gives  ground-state energies that smoothly change with carrier concentration.
By using $E_0^{\boldsymbol{\kappa}}$, the Drude weight $D$ along the $x$ direction is given by 
$D=N_\kappa^{-1}\sum_{\boldsymbol{\kappa}} (2N)^{-1} \partial^2E_0^{\boldsymbol{\kappa}}/\partial \kappa_x^2$.

\begin{figure}
\begin{center}
\includegraphics[width=6.cm]{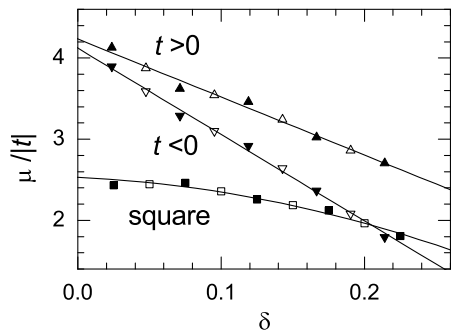}
%\vskip4cm
\caption{\label{fig2}
Chemical potential $\mu$ versus $\delta$ in the strong-coupling limit of the Hubbard model with $U/t=20$. The upper and lower triangles represent the data for the $N=21$ triangular lattice with $t>0$ and $t<0$, respectively, and the squares are for the $N=20$ square lattice. The filled and open symbols correspond to $\mu$ obtained by taking the concentration difference of $\delta_2-\delta_1=1/N$ and $2/N$, respectively. The solid lines are guides for eyes with linear and quadratic functions for the triangular and square lattices, respectively.}
%\vspace{6mm}
\end{center}
\end{figure}

In Fig.~\ref{fig2}, $\mu$ is plotted as a function of $\delta$. $\mu$ for both the $t>0$ and $t<0$ triangular lattices follows well the linear solid lines, implying that $\mu$ is proportional to $\delta$.  On the other hand, $\mu$ in the square lattice clearly deviates from a linear relation, but is well described by a quadratic polynomial. The relation that $\mu\propto\delta^2$ is consistent with previous numerical results.~\cite{Furukawa,Kohno} This contrasting behavior between the triangular and square lattices indicates essential difference of their electronic states.

According to quantum criticality hypothesis of the MI transition, a hyper scaling theory has been developed.~\cite{Imada}  The dynamical critical exponent $z$ is related to the doping dependence of $\mu$ in the form that $\mu-\mu_c\propto \delta^{z/d}$, where $d$ is the space dimension and $\mu_c$ is the chemical potential at a critical concentration ($\delta=0$). The relation that $z=2$ is obtained for the 1D Hubbard model from the exact solution.~\cite{Lieb} From Fig.~\ref{fig2}, the $d=2$ triangular lattice also satisfies $z=2$. However, in the square lattice, $z=4$ since $\mu-\mu_c\propto\delta^2$.~\cite{Imada}

\begin{figure}
\begin{center}
\includegraphics[width=8.cm]{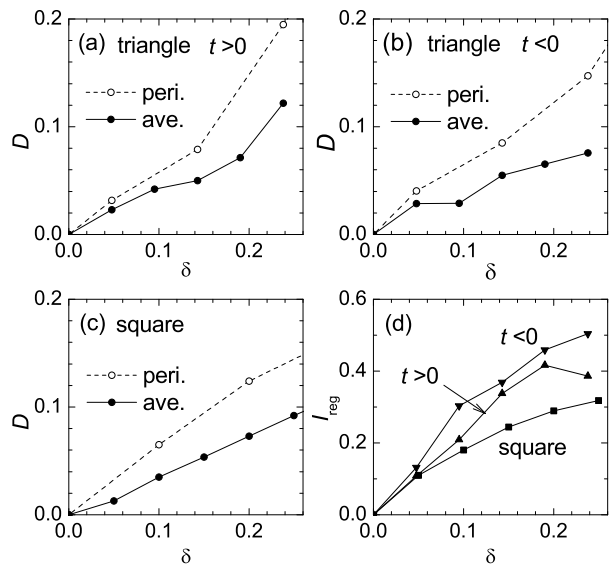}
%\vskip4cm
\caption{\label{fig3}
Drude weight $D$ versus $\delta$ in the strong-coupling limit of the Hubbard model with $U/t=20$ for the $N=21$ triangular lattice with (a) $t>0$, (b) $t<0$, and (c) the $N=20$ square lattice. The solid circles are obtained by averaging over twisted BC. The open circles represent the data under periodic BC and for the even number of electrons. (d) Doping dependence of the integrated weight of $\sigma_\mathrm{reg}(\omega)$. The upper and lower triangles represent the data for the $N=21$ triangular lattice with $t>0$ and $t<0$, respectively, and the squares are for the $N=20$ square lattice.}
%\vspace{6mm}
\end{center}
\end{figure}

The hyperscaling theory also predicts that $D\propto \delta^{1+(z-2)/d}$.~\cite{Imada} Since $z=2$ for the triangular lattices, we expect that $D\propto \delta$. Figures ~\ref{fig3}(a) and \ref{fig3}(b) show the doping dependece of $D$. The open and solid circles represent $D$ obtained for the periodic BC and for the averaging procedure, respectively. Although $D$ does not necessarily show smooth behaviors because of the finite-size effect, it seems to grow roughly proportional to $\delta$ with a slight convex behavior below $\delta\sim 0.2$, i.e., $D\propto \delta^\alpha$ with $\alpha\leqslant 1$, for both the $t>0$ and $t<0$ cases. It is difficult to estimate the value of $\alpha$ from these data, but we consider that the data do not contradict the predicted behavior of $D\propto\delta$ from $\mu$. This suggests that the scaling theory is satisfied in the triangular Hubbard model. Since the scaling relation with $z=2$ is also satisfied in the 1D Hubbard model where charge is decoupled from spin in the ground state, we can say that, even in the triangular lattice, charge would be coupled very weakly to spin. This view is consistent with the results of spin correlation discussed above.

In the square lattice, two different behaviors have been reported: $D$ is proportional to $\delta^2$ (Ref.~21) and $\delta$ (Refs.~22,23).   Figure~\ref{fig3}(c) shows proportionality to $\delta$. The resulting $z=2$ is different from $z=4$ deduced from $\mu$. This inconsistency may tell us that the MI transition of the square Hubbard lattice is not a continuous phase transition.  This is in contrast with  the 1D case and probably with the triangular lattice. The difference might be attributed to the presence of strong coupling of charge and spin in the square lattice.  However, we have to say that our numerical data is still not enough to give a final conclusion, because $D$ very close to half filling is not rigorously linear in $\delta$. Therefore, we need further careful efforts to solve this important problem.

The real part of the optical conductivity reads
$\sigma(\omega)=2\pi D \delta(\omega)+\sigma_\mathrm{reg}(\omega)$,
where the regular part $\sigma_\mathrm{reg}(\omega)$ under the averaging procedure is given by
\begin{equation*}
\sigma_\mathrm{reg}(\omega)=\frac{1}{N_\kappa}\sum_{\boldsymbol{\kappa}} \frac{\pi}{N\omega} \sum_m \left|\left< \Psi^{\boldsymbol{\kappa}}_m \left| j_x^{\boldsymbol{\kappa}} \right|\Psi^{\boldsymbol{\kappa}}_0 \right>\right|^2 \delta (\omega- E^{\boldsymbol{\kappa}}_m + E^{\boldsymbol{\kappa}}_0),
\end{equation*}
where $\left| \Psi^{\boldsymbol{\kappa}}_m\right\rangle$ represents an eigenstate with energy $E^{\boldsymbol{\kappa}}_m$ for a given $\boldsymbol{\kappa}$.  
The $x$ component of the current operator is given by $j_x=\mathrm{i}\left[H_\mathrm{SC}, \hat{x}\right]$,
where $\hat{x}$ is the $x$ component of the total position operator.
A standard continued-fraction expansion method based on the Lanczos algorithm  is used to calculate $\sigma_\mathrm{reg}(\omega)$.

\begin{figure}
\begin{center}
\includegraphics[width=6.cm]{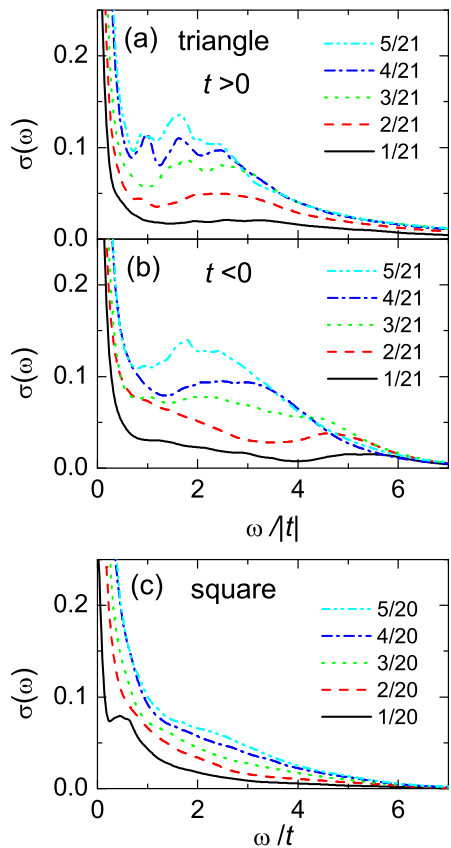}
%\vskip4cm
\caption{\label{fig4}
(Color online) Optical conductivity $\sigma(\omega)$ for various hole concentrations of the strong-coupling limit of the Hubbard model with $U/t=20$. (a) $t>0$, (b) $t<0$ for the $N=21$ triangular lattice, and (c) the $N=20$ square lattice. The data are obtained with the Lorentzian broadening of $0.1t$. The numbers in the panels represent the hole concentration.}
%\vspace{6mm}
\end{center}
\end{figure}

Figure~\ref{fig4} shows the doping dependence of $\sigma(\omega)$. In the triangular lattices, we find that incoherent excitations are widely extended to a high-energy region up to several $t$. This sounds counterintuitive from the viewpoint of weak coupling of charge to spin in the ground state mentioned above,
since, in the 1D Hubbard model where spin and charge are completely decoupled, incoherent weights are negligibly small.~\cite{Horsch}  
The incoherent excitations are also different from those of the square lattice. For the square lattice, the origin of the incoherent part is well-understood based on the picture that magnetic polarons move in the system by inducing magnetic excitations scaled by the exchange interaction $J=4t^2/U\sim 0.2t$. Therefore, the incoherent spectral weight shown in Fig.~\ref{fig3}(c) is mainly located below $\omega\sim t$. Very incoherent $\sigma(\omega)$ up to several $t$ in the triangular lattices clearly indicates that such a magnetic-polaron picture cannot be applied.

An interesting feature of $\sigma(\omega)$ for both the $t>0$ and $t<0$ triangular lattices is that the incoherent spectral weight increases with increasing $\delta$ without changing their global energy distribution in the region of $t\lesssim\omega\lesssim 5t$. Furthermore, the spectral weight of the incoherent part is larger than that of the square lattice, as shown in Fig.~\ref{fig3}(d) where the integrated weight of $\sigma_\mathrm{reg}(\omega)$  is plotted. Since the magnitude of $D$ is comparable between the triangular and square lattices, the large incoherent weight in the triangular lattice means that the motion of carrier is more incoherent than the square lattice. Enhanced incoherent structures have also discussed in terms of the single-hole spectral function in triangular antiferromagnet.~\cite{Trumper} Both quantities may have common origins.
A possible origin is nontrivial phases from the spin degrees of freedom~\cite{Weng} with strong frustration. The phases may act as random fields on carriers and induce incoherent excitations. Obtaining analytical expressions for theses effects is a future problem.
We note that such an enhanced incoherency might be a possible origin of anomalous incoherent contribution of $\sigma(\omega)$ observed in a triangular compound Na$_{0.7}$CoO$_2$.~\cite{Wang}

In summary, we have performed the exact diagonalization study of the doping dependence of the optical conductivity $\sigma(\omega)$, the chemical potential $\mu$, and static spin correlation for the triangular Hubbard model. We find that $\mu$ and the Drude weight $D$ near half filling are roughly proportional to the hole concentration. These behaviors are similar to those of the 1D Hubbard model where the charge degree of freedom is decoupled from spin, but different from the case of the square lattice where the coupling is expected to be strong.  Therefore, charge in the triangular lattice is expected to be weakly coupled to spin. Spin correlation function near half filling supports this view. However, we find that $\sigma(\omega)$ shows very strong incoherent excitations, which are absent in the square lattice. This implies that geometrical frustration influences incoherent charge dynamics. A clear understanding of the physics behind this remains for the future. 

I would like to thank W. Koshibae, K. Tsutsui, N. Bulut, S. Maekawa and H. Fukuyama for useful discussions.  This work was supported by the Next Generation Super Computing Project of Nanoscience Program, CREST, and Grant-in-Aid for Scientific Research from MEXT.  The numerical calculations were partly performed in the supercomputing facilities in ISSP, University of Tokyo and in IMR, Tohoku University.

\end{document}